\documentclass{ctr}

\usepackage{ctrfont}
\usepackage{natbib}

\usepackage{url}
\usepackage{amsmath}
\usepackage{amssymb}
\usepackage[symbol]{footmisc}

\usepackage{graphicx}

\usepackage{psfrag}
\usepackage{subfig}

\usepackage[labelsep=period]{caption}
\renewcommand{\tablename}{\sc Table}
\renewcommand{\figurename}{\sc Figure}

\title{Self-critical machine-learning wall-modeled LES for external aerodynamics}

\shorttitle{Self-critical machine-learning model for LES}

\author{A. Lozano-Dur\'an \and H.~J. Bae\footnote{Harvard University,
    Cambridge, MA 02138, USA}}

\shortauthor{Lozano-Dur\'an \& Bae}

\begin{document}
\pagenumbering{gobble}


\maketitle


\section{Motivation and objectives}

The use of computational fluid dynamics (CFD) for external aerodynamic
applications has been a key tool for aircraft design in the modern
aerospace industry. However, flow predictions from the
state-of-the-art solvers are still unable to comply with the stringent
accuracy requirements and computational efficiency demanded by the
industry. In recent years, wall-modeled large-eddy simulation (WMLES)
has gained momentum as a high-fidelity tool for routine industrial
design. In WMLES, only the large-scale motions in the outer region of
the boundary layer are resolved, which enables a competitive
computational cost compared with other CFD
approaches~\citep{Choi2012}. As such, NASA has recognized WMLES as an
important pacing item for ``developing a visionary CFD capability
required by the notional year 2030''~\citep{Slotnick2014}. In the
present brief, we introduce a wall model based on the flow-state
classification that is also self-critical, i.e., it provides a
confidence value for the classification.

Several strategies for modeling the near-wall region have been
explored in the literature, and comprehensive reviews can be found in
\citet{Cabot2000}, \citet{Piomelli2002}, \citet{Spalart2009},
\citet{Larsson2015}, and \citet{Bose2018}. One of the most widely used
approaches for wall modeling is the wall-flux modeling approach (or
approximate boundary conditions modeling), where the no-slip and
thermal wall boundary conditions are replaced with stress and
heat-flux boundary conditions provided by the wall model. This
category of wall models utilizes the large-eddy simulation (LES)
solution at a given location in the LES domain as input and returns
the wall fluxes needed by the LES solver. Examples of the most popular
and well-known approaches are those computing the wall stress using
either the law of the wall \citep{Deardorff1970, Schumann1975,
  Piomelli1989}, the full/simplified RANS equations
\citep{Balaras1996, Wang2002, Chung2009, Bodart2011, Kawai2013,
  Bermejo-Moreno2014, Park2014, Yang2015} or dynamic wall
models~\citep{Bose2014, Bae2019}.

In recent years, advances in machine learning and data science have
incited new efforts to complement the existing turbulence modeling
approaches in the fluids community. However, machine learning is still
far from being the panacea to solve long-standing problems in LES.
The statement ``LES modeling will be solved by machine learning'' is
as meaningful as ``LES modeling will be solved by the Fourier
transform'', i.e., not very. Machine learning is a tool, as much as the
Fourier transform is a tool. Both might aid the modeling of turbulent
flows if properly used.  Ultimately, in-depth knowledge of the physics
to be modeled and the formulation of the problem in a language
consistent with machine learning is key to utilizing the tool to its
fullest potential.

Supervised learning, i.e., the machine-learning task of learning a
function that maps an input to an output based on provided training
input-output pairs, was first introduced in turbulence modeling in the
form of subgrid-scale (SGS) modeling for LES. Early approaches used
neural networks to emulate and speed up a conventional, but
computationally expensive, SGS model~\citep{Sarghini2003}. More
recently, SGS models have been trained to predict the (so-called)
perfect SGS terms using data from filtered direct numerical simulation
(DNS) \citep{Gamahara2017,Xie2019}.  Other approaches include deriving
SGS terms from optimal estimator theory~\citep{Vollant2017} and
deconvolution operators~\citep{Hickel2004,Maulik2017,Fukami2019}.  One
of the first attempts at using supervised learning for WMLES can be
found in \citet{Yang2019}. These authors noted that a model trained on
channel flow data at a single Reynolds number could be extrapolated to
higher Reynolds numbers and similar configurations. The model relied
on information about the flow that is typically inaccessible in
real-world applications, such as the boundary-layer thickness, and was
limited to channel flow configurations. The reader is referred to
\cite{Duraisamy2019} and \cite{Brunton2019} for a literature review on
machine learning for fluid mechanics.

Currently, the major challenge for WMLES of realistic external
aerodynamic applications is achieving the robustness and accuracy
necessary to model the myriad of different flow regimes that are
characteristic of these problems. Examples include turbulence with
mean-flow three-dimensionality, laminar-to-turbulent transition, flow
separation, secondary flow motions at corners, and shock wave
formation, to name a few. The wall-stress generation mechanisms in
these complex scenarios differ from those in flat plate turbulence.
However, the most widespread wall models are built upon the assumption
of statistically-in-equilibrium wall-bounded turbulence, which only
applies to a handful number of flows. The latter raises the question
of how to devise models capable of seamlessly accounting for such a
vast and rich collection of flow physics in a single unified approach.

In the present brief, we develop a wall-flux-based wall model for LES
using a self-critical machine-learning approach. Since data-driven
models are limited by the information they are provided, the model is
formulated to naturally account for various flow configurations. In
this preliminary work, the wall model is trained on DNS data of flow
over a flat plate, flow in a turbulent duct, and separated flow at
various Reynolds numbers. The model comprises two components: a
classifier and a predictor. The classifier is trained to place the
flows into the separate categories along with a confidence value,
while the predictor outputs the modeled wall stress based on the
likelihood of each category.  The model is validated on a flow over
the fuselage and wing-body junction of the NASA Juncture Flow
Experiment.

This brief is organized as follows. The flow setup, mathematical
modeling, and numerical approach are presented in Section
\ref{sec:methods}.  Results for WMLES with a widely used equilibrium
wall model are presented in Section~\ref{sec:results_EQWM}. The
results include the prediction of the mean velocity profiles for three
different locations on the aircraft: the upstream region of the
fuselage, the wing-body juncture, and wing-body juncture close to the
trailing edge.  The formulation of the new model is discussed in
Section \ref{sec:formulation}. The model is validated in Section
\ref{sec:validation} and compared with the equilibrium wall
model. Finally, conclusions are offered in Section
\ref{sec:conclusions}.

\newpage

\section{Simulation setup and methods}\label{sec:methods}

\subsection{NASA Juncture Flow}

\begin{figure}
  \centering
  \vspace{0.5cm}
\includegraphics[width=0.9\textwidth]{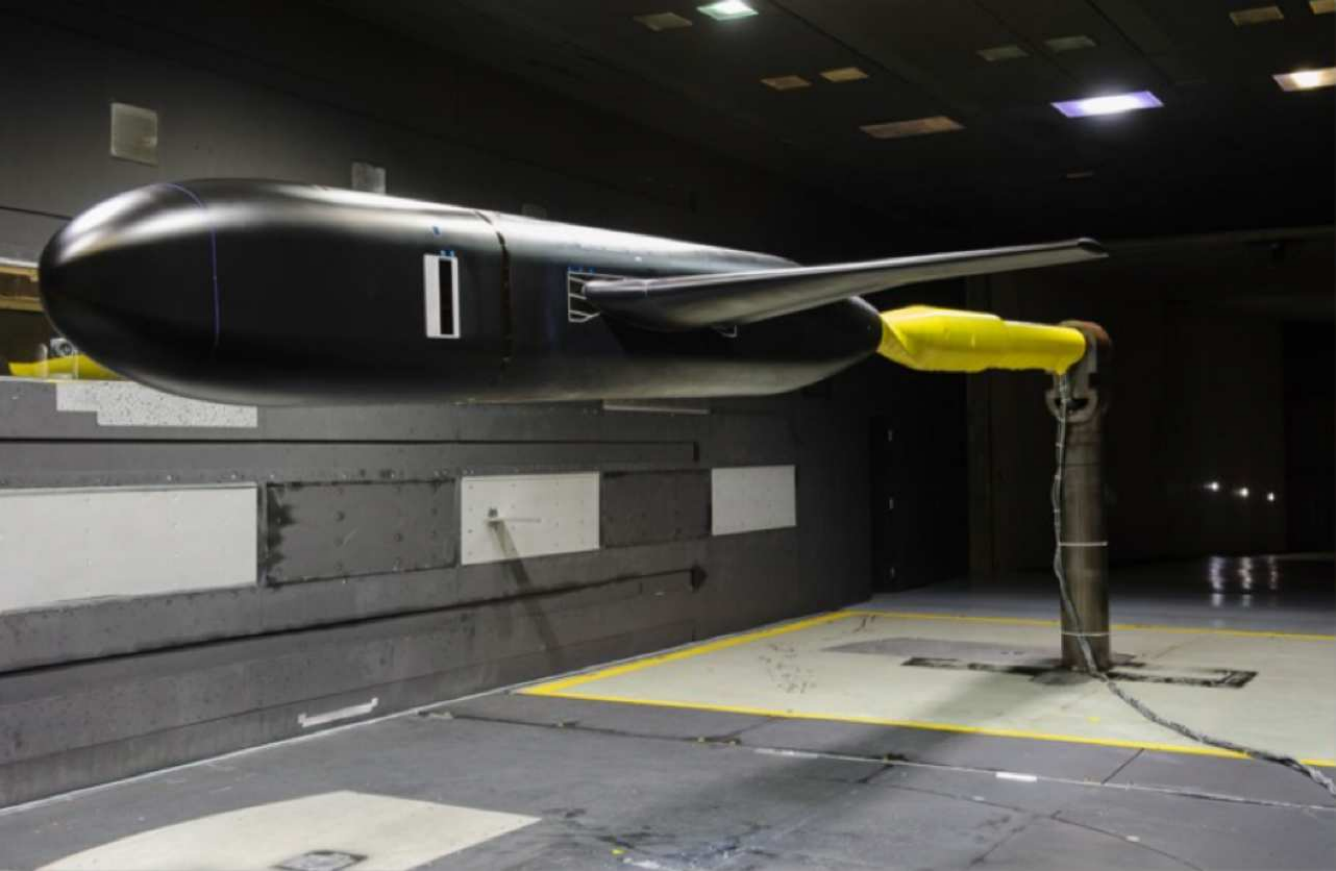}
\caption{Experimental setup of the NASA Juncture Flow at NASA Langley
  14- by 22-foot Subsonic Wind Tunnel.\label{fig:experiment}}
\end{figure}

The problem considered is the NASA Juncture Flow, which has been
recently proposed as a validation experiment for generic wing-fuselage
junctions at subsonic conditions~\citep{Rumsey2019}. The experiment
consists of a full-span wing-fuselage body configured with truncated
DLR-F6 wings and has been tested in the Langley 14- by 22-foot
Subsonic Tunnel (Figure \ref{fig:experiment}). The Reynolds number is
$Re=L U_\infty/\nu=2.4$ million, where $L$ is the chord at the Yehudi
break, $U_\infty$ is the freestream velocity and $\nu$ is the
kinematic viscosity. The density of the air is $\rho$. We consider an
angle of attack of $AoA=5$ degrees. The experimental dataset comprises
a collection of local-in-space time-averaged
measurements~\citep{Kegerise2019}, such as velocity profiles and
Reynolds stresses, which greatly aid the validation of models and their
ability to capture the critical flow physics. The frame of reference
is such that the fuselage nose is located at $x = 0$, the $x$-axis is
aligned with the fuselage centerline, the $y$-axis denotes the spanwise
direction, and the $z$-axis is the vertical direction.  The associated
instantaneous velocities are denoted by $u$, $v$, and $w$, and
occasionally by $u_1$, $u_2$, and $u_3$. Time-averaged quantities are
denoted by $\langle \cdot \rangle$.
%

\subsection{Numerical methods and models}

\begin{figure}
  \centering
    \vspace{0.5cm}
\includegraphics[width=1\textwidth]{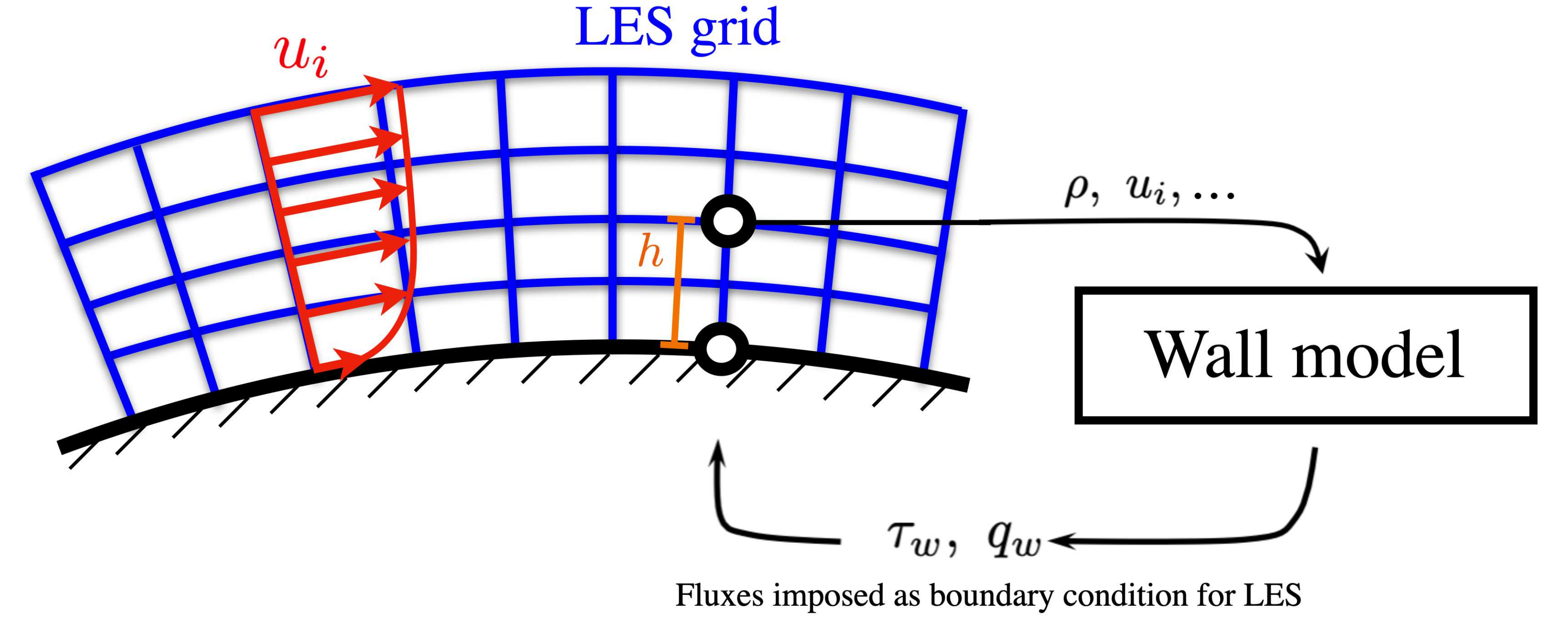}
\caption{Schematic of the working principle of wall-modeled large-eddy
  simulation.\label{fig:WMLES}}
\end{figure}

We perform WMLES of the NASA Juncture Flow using the high-fidelity
solver charLES. The reader is referred to \cite{Lozano_brief_2020_1}
for a detailed description of the numerical methods, grid generation
and modeling approach. Here, we provide an abridged description of the
numerical setup. The code integrates the compressible LES equations
using a kinetic-energy conserving, second-order accurate, finite
volume method. The SGS model is the dynamic Smagorinsky model
\citep{Germano1991} with the modification by \cite{Lilly1992}.  We
utilize a wall model to overcome the restrictive grid-resolution
requirements to resolve the small-scale flow motions in the vicinity
of the walls. The working principle of WMLES is illustrated in Figure
\ref{fig:WMLES}. Flow information from the LES grid at a wall-normal
distance $h$ is used as input to the wall model. The wall model
predicts the wall stress $\tau_w$ and heat flux at the wall $q_w$,
which are imposed back to LES as boundary conditions.  We use an
algebraic equilibrium wall model (EQWM) derived from the integration
along the wall-normal direction of an assumed constant-stress
layer~\citep{Wang2002, Kawai2012, Larsson2015}. At each point of the
wall $\boldsymbol{x}$, this yields an equation for the wall-model mean
velocity profile
\begin{equation} \label{eq:algwm}
  {u}_{||}^+(y_n^+) =
  \begin{cases}
    y_n^+ + a_1 (y_n^+)^2 & \text{for $y_n^+ < y_n^\mathrm{ref}$}, \\
    \frac{1}{\kappa}\ln(y_n^+) + B & \text{otherwise,}
  \end{cases}
\end{equation}
where ${u}_{||}$ is the magnitude of the wall-parallel velocity, $y_n$
is the wall-normal coordinate such that $y_n=0$ is the wall, $+$
denotes normalization by $\nu$ and $\tau_w$, $\kappa$ = 0.41, $B$ =
5.2, $y_n^\mathrm{ref}$ $=$ 23, and $a_1\approx-0.02$. The wall stress
$\tau_w$ is computed by solving Eq.  (\ref{eq:algwm}) iteratively
evaluated at the wall-parallel LES velocity $u_{||}(\boldsymbol{x} +
\boldsymbol{e}_2 h)$, where $\boldsymbol{e}_2$ is the wall-normal
direction. The matching location $h$ for the wall model is the first
off-wall cell center of the LES grid. The walls are assumed
isothermal. The modeling parameters in Eq.~(\ref{eq:algwm}) are
calibrated to match the wall-stress prediction for a
zero-pressure-gradient turbulent boundary layer (ZPGTBL).
%

\subsection{Grid generation}

\begin{figure}
    \begin{center}
      \includegraphics[width=1.0\textwidth]{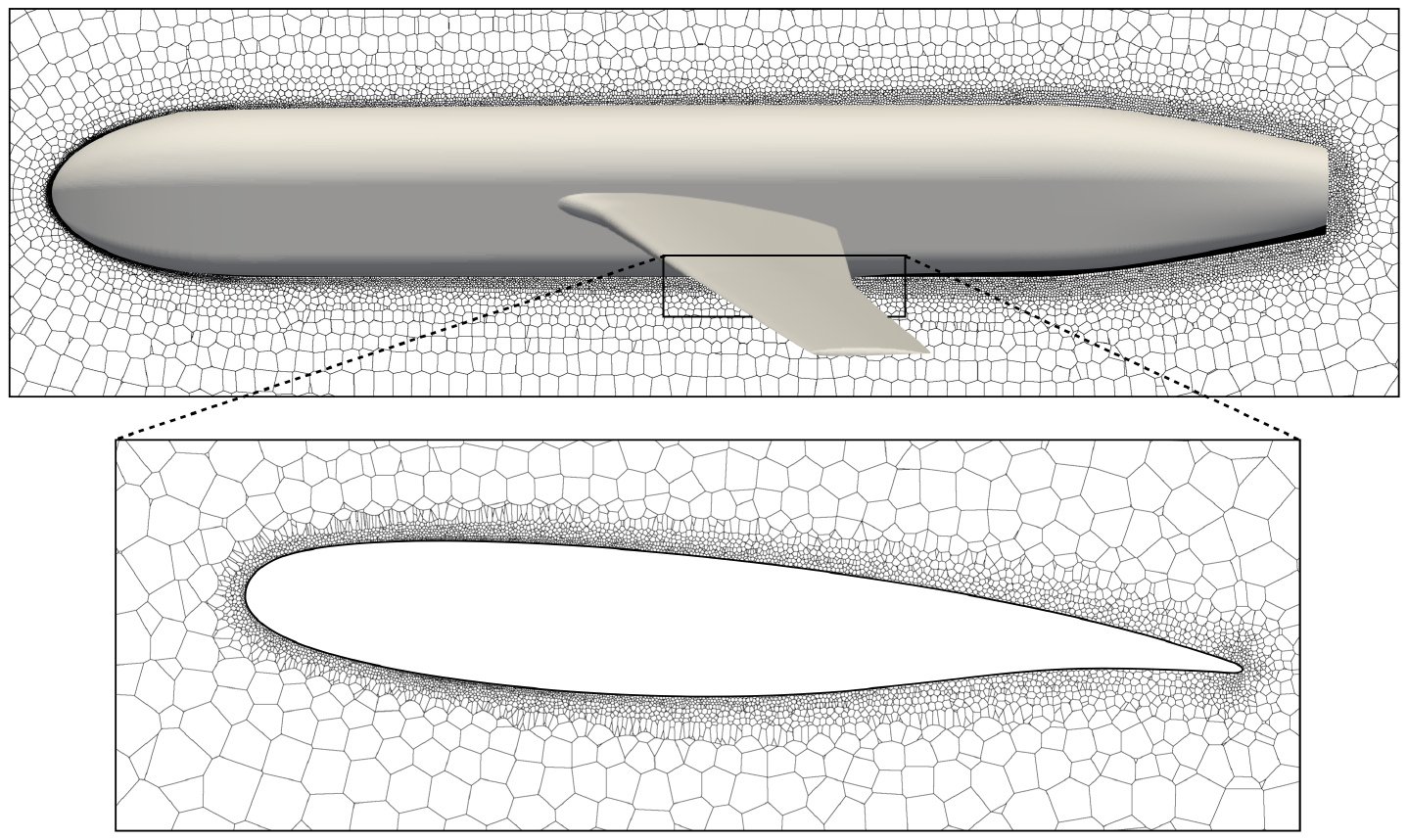}
  \end{center}
  \caption{Visualization of Voronoi control volumes for
    boundary-layer-conforming grid with $N_{bl} = 5$ and
    $Re_\Delta^\mathrm{min}=8 \times 10^3$~\citep[see][for
      details]{Lozano_brief_2020_1}.\label{fig:tbl_grids}}
\end{figure}

\begin{figure}
\centering
\includegraphics[width=1\textwidth]{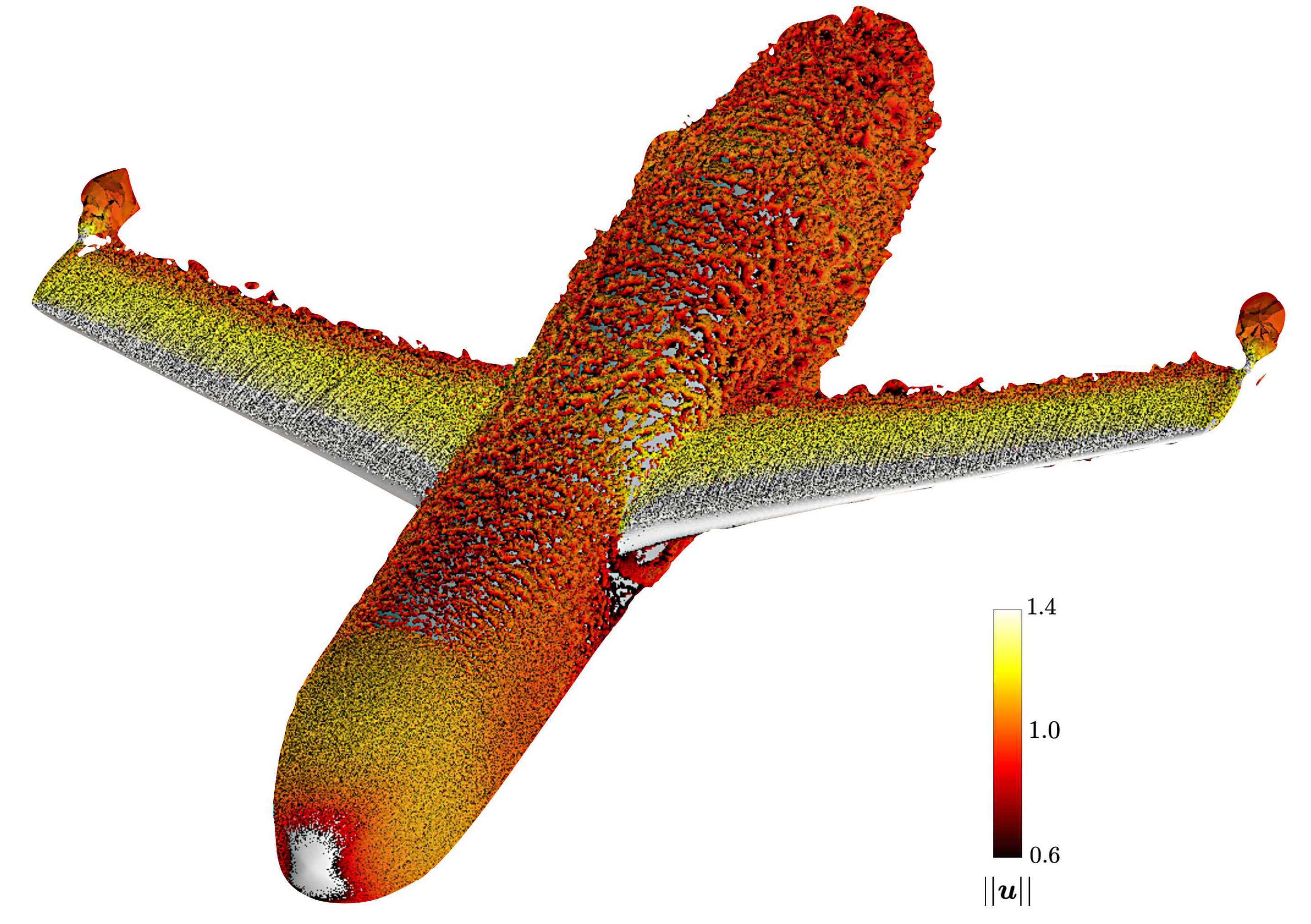}
\caption{Visualization of the instantaneous isosurfaces of the
  Q-criterion colored by the velocity magnitude.\label{fig:Snap}}
\end{figure}

The mesh generation is based on the Voronoi tessellation of a
collection of points.  The points are seeded using a
boundary-layer-conforming strategy such that the number of points per
boundary-layer thickness is constant and equal to $N_{bl}=5$.  The
minimum grid Reynolds number resolved is $Re_{\Delta}^\mathrm{min}
\equiv \Delta_\mathrm{min} U_\infty/\nu =8\times 10^3$, where
$\Delta_\mathrm{min}$ is the smallest grid size allowed. Figure
\ref{fig:tbl_grids} shows two cuts of the grid along the fuselage and
wing, and Figure \ref{fig:Snap} contains an isosurface of the
instantaneous Q-criterion.
%
%

\section{Results of WMLES with EQWM}\label{sec:results_EQWM}

\begin{figure}
  \centering
  \vspace{0.5cm}
\includegraphics[width=1\textwidth]{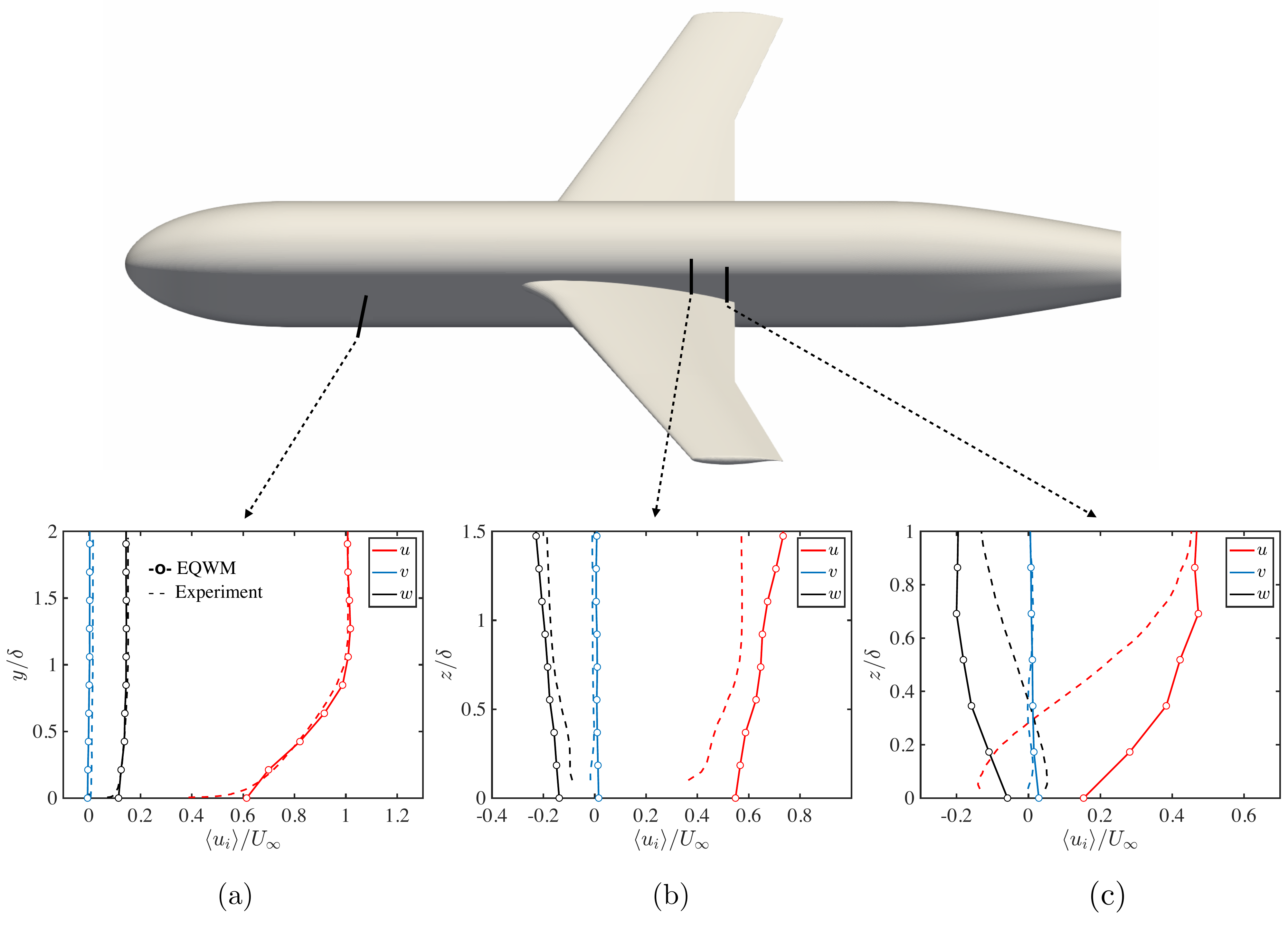}
\caption{ The mean velocity profiles at (a) upstream region of the
  fuselage at $x=1168.4$~mm and $z=0$~mm, (b) wing-body juncture at
  $x=2747.6$~mm and $y=239.1$~mm, and (c) wing-body juncture close to
  the trailing edge at $x=2922.6$~mm and $y=239.1$~mm. Solid lines with
  symbols denote WMLES with EQWM and dashed lines are
  experiments. Colors denote different velocity components. The
  distances $y$ and $z$ are normalized by the local boundary-layer
  thickness $\delta$ at each location.\label{fig:results_EQWM}}
\end{figure}

The WMLES discussed above constitutes our baseline case. For
reference, this case corresponds to case C-N5-Rem8e3 in
\citet{Lozano_brief_2020_1}.  The prediction of the mean velocity
profiles is shown in Figure~\ref{fig:results_EQWM} and compared with
experimental measurements ($\boldsymbol{u}^{\mathrm{exp}}$).  Three
locations are considered: the upstream region of the fuselage, the
wing-body juncture\footnote{Note that this location differs from the
  wing-body juncture location selected in \cite{Lozano_brief_2020_1}.
  The current location was chosen to highlight the deficiencies of the
  EQWM.}, and the wing-body juncture close to the trailing edge.

In the first region, the flow resembles a ZPGTBL. Hence, the dynamic
Smagorinsky SGS model and the equilibrium wall-model in
Eq.~(\ref{eq:algwm}) perform accordingly (i.e., errors below 2\%), as
these have been devised for and validated in ZPGTBL. On the contrary,
there is a decline of accuracy in the WMLES results in the wing-body
juncture and trailing-edge region, which are dominated by secondary
motions in the corner and flow separation. \cite{Lozano_brief_2020_1}
have further shown that not only are the errors larger in the
wing-body juncture and trailing-edge region, but the rate of
convergence of WMLES by merely refining the grid is too slow to
compensate for the modeling deficiencies.
%

The results above suggest that novel modeling venues must be exploited
to improve WMLES predictions at an affordable computational cost.
Here, we focus our efforts on wall model improvements. Nonetheless, we
remark that physical insights, novel SGS modeling, and
numerical/gridding advancements are also essential to attain robust
high-accuracy WMLES. Future work will be devoted to devising an
integrated modeling approach.
 
\section{Model formulation}\label{sec:formulation}

\begin{figure}
      \centering
      \includegraphics[width=1\textwidth]{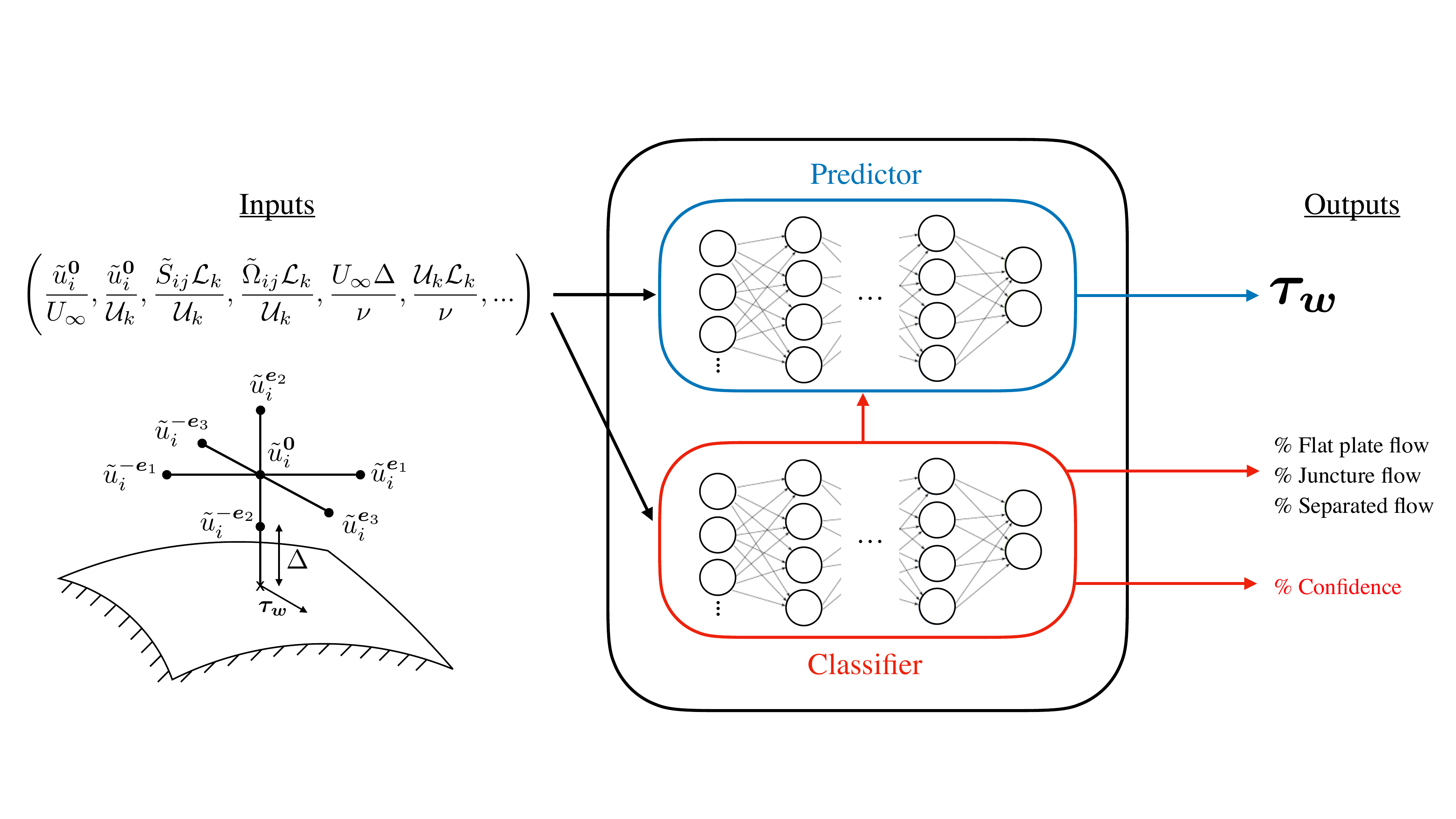}
      \caption{Schematic of the self-critical model (SCM). Details are
        provided in Section~\ref{subsec:IO}.
      \label{fig:model_sketch}}
\end{figure}

The working principle of the proposed model is summarized in
Figure~\ref{fig:model_sketch}. The self-critical model (SCM) is
comprised of two elements: a classifier and a predictor.  First, the
classifier is fed with data from the LES solver and quantifies the
similarities of the input with a collection of known building-block
flows. The predictor leverages the information of the classifier to
generate a wall-stress prediction via non-linear interpolation of the
building-block database. The model is self-critical, i.e., it also
outputs a high confidence value to the prediction if the input flow is
recognized as a combination of the building blocks.  If the input data
looks extraneous, the model provides a low confidence value, which
essentially means that the flow is unknown.  We refer to the current
version of the model as the SCM version 1 (SCMv1).

\subsection{Model requirements and assumptions}

We consider three basic model requirements. First, to comply with
dimensional consistency, the inputs and outputs of the model should be
given in non-dimensional form. Second, the model should be invariant
under constant space/time translations and rotations of the frame of
reference. Finally, the model should satisfy Galilean invariance,
i.e., invariant under uniform velocity transformations of the frame of
reference (Galilean invariance) or unidirectional accelerations in the
case of incompressible flows (extended Galilean invariance).

\begin{figure}
  \centering
  \includegraphics[width=1\textwidth]{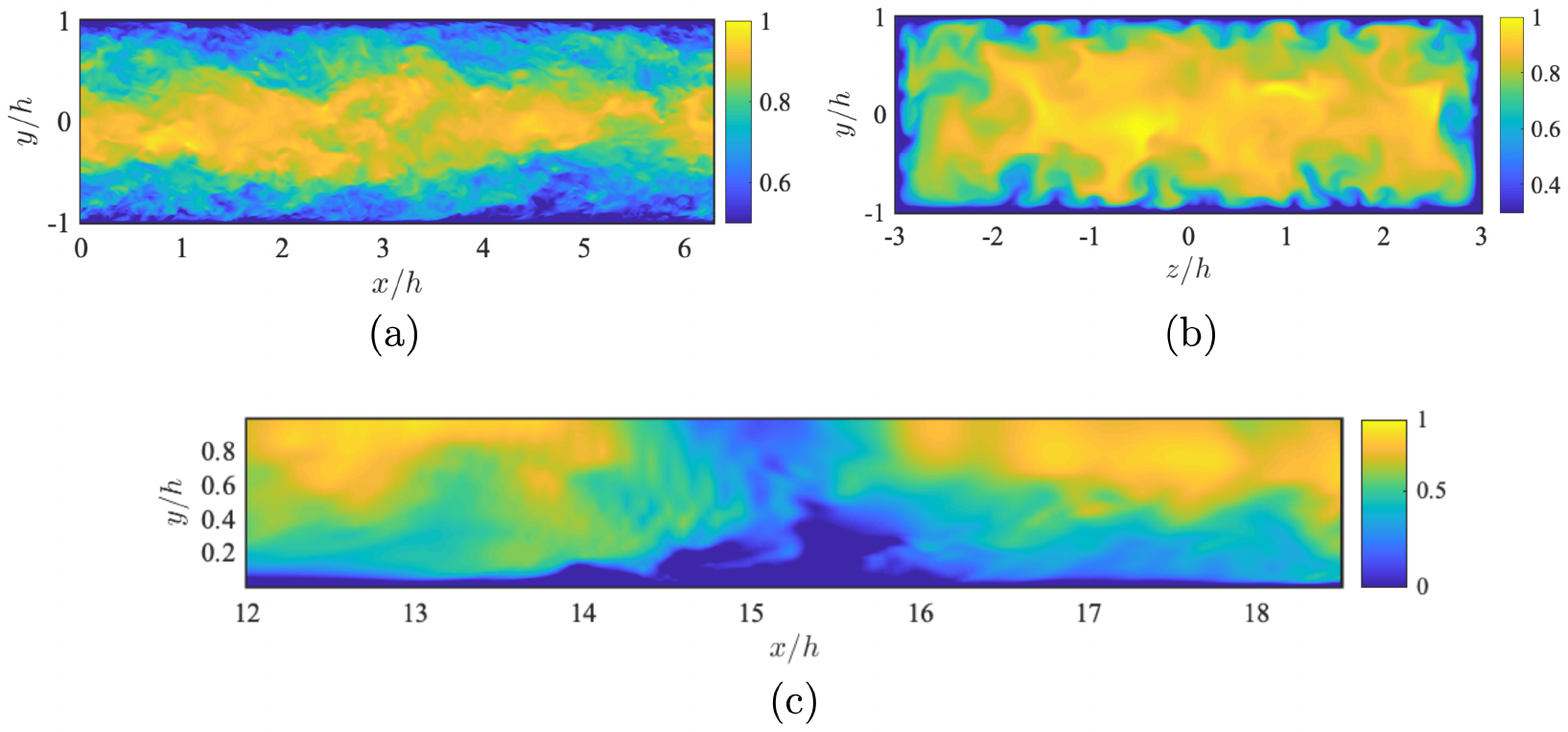}
  \caption{Three examples of the building block units taken as
    representative of the potential flow configurations. (a) Turbulent
    channel flow at $Re_\tau=1000$, (b) turbulent duct at
    $Re_\tau=180$ and aspect ratio 3, and (c) boundary layers with
    a separation bubble via blowing and suction at $Re_\tau\approx
    180$. The colors are the instantaneous streamwise velocity
    normalized by the freestream/centerline velocity. Panels (a) and
    (c) show the velocity for a streamwise/wall-normal plane and panel
    (b) for a wall-normal/spanwise plane. \label{fig:flow_units}}
\end{figure}

The main model assumption is that the myriad of flow configurations
encountered in external aerodynamics might be represented by a finite
(hopefully small) set of canonical flow units. The assumption relies
on the idea that a collection of building blocks contains the
essential flow physics necessary to formulate generalizable models.
In this preliminary work, the set of canonical units selected
includes: turbulent channel flows, turbulent ducts, and turbulent
boundary layers with separation. The three flow units are
representative of ZPGTBL, turbulent flow in junctures, and separated
turbulence, respectively.  Examples of the three building-block units
are included in Figure \ref{fig:flow_units}.  The use of merely three
building blocks is far from being representative of the rich flow
physics (i.e., laminar flows, shock waves, compressibility effects,
flow unsteadiness, adverse/favorable pressure gradients, other
mean-flow three-dimensionalities and separation patterns, chemical
reactions, etc.). Here, we test the model trained with only three
building blocks and the collection will be extended in upcoming
versions.
%

\subsection{Input and output variables}\label{subsec:IO}

The input velocity is acquired using a seven-point stencil as shown in
Figure \ref{fig:model_sketch}.  The center of the stencil,
$\boldsymbol{x}^{\boldsymbol{0}}$ is located at a wall-normal distance
$h$ away from the wall, which corresponds to the second grid point off
the wall, i.e., $h\approx 2\Delta$. The other six components of the
stencil are located $\Delta$ away from
$\boldsymbol{x}^{\boldsymbol{0}}$ and are denoted as
$\boldsymbol{x}^{\pm\boldsymbol{e}_i}= \boldsymbol{x}^{\boldsymbol{0}}
\pm \Delta \boldsymbol{e}_i$ for $i=1,2,3$, where $\boldsymbol{e}_i$
is a unit vector. The vector $\boldsymbol{e}_2$ is parallel to the
wall-normal direction and $\boldsymbol{e}_1$ and $\boldsymbol{e}_3$
are oriented to form an orthogonal basis. The velocity components at
the stencil points are denoted by $u_i^{\boldsymbol{0}}$ and
$u_i^{\pm\boldsymbol{e}_i}$. The input variables are arranged as
velocity differences
\begin{subequations}\
\begin{gather}
  \tilde{u}_i^{\boldsymbol{0}} = u_i^{\boldsymbol{0}}-u_{i,\mathrm{wall}}, \\
  \tilde{S}_{ij} = (\delta_iu_j+\delta_ju_i)/2,  \\
  \tilde{\Omega}_{ij} = (\delta_iu_j-\delta_ju_i)/2,
\end{gather}
\end{subequations}
where $\delta_iu_j = (u_j^{\boldsymbol{e}_i} -
u_j^{-\boldsymbol{e}_i})/(2\Delta)$ and $u_{i,\mathrm{wall}}$ is the
velocity at the wall. The velocities $u_i$ are obtained by
interpolation of the LES velocities to
$\boldsymbol{x}^{\boldsymbol{0}}$ and
$\boldsymbol{x}^{\pm\boldsymbol{e}_i}$.

The multi-space structure of a seven-point stencil offers important
advantages over the traditional single-point stencil. A key property
is that it makes possible to discern among different types of flow,
such as separation zones, mean-flow three-dimensionality, or even
whether the input variable is within the boundary layer or the
freestream. This capability is unavailable using single-point
stencils. The flow-state variables $\tilde{u}_i^{\boldsymbol{0}}$,
$\tilde{S}_{ij}$ and $\tilde{\Omega}_{ij}$ comply with Galilean
invariance, as they are constructed using relative velocities. Note
that $\delta_i$ differs from the discrete gradient operators in
charLES and the stencil points do not necessarily coincide with the
grid cell centers. In unstructured grids, the location of the cell
centers could follow a complex pattern in space. Additionally, the
discrete gradient operators may differ depending on the discretization
method. Hence, the stencil selected aims at alleviating these problems by
facilitating the model training process from DNS data and promoting
generalizability for different numeric and grid strategies.
 
In addition to velocity information, the model also takes into account
the grid size $\Delta$, the distance to the wall $h$, the fluid
properties $\rho$ and $\nu$, and the farfield velocity $U_\infty$
(defined with respect to the wall velocity). To ease the
classification of different flow regimes, we introduce a
characteristic length-scale $\mathcal{L}$ and velocity scale
$\mathcal{U}$ for each of the building blocks considered. In the
turbulent channel, the characteristic scales are assumed to be the
stress at the wall and the distance to the wall, which yield
$\mathcal{L}_1 = h$ and
$\mathcal{U}_1=\sqrt{\tau_w/\rho}$~\citep{Townsend1976}. In the steady
turbulence with mean-flow three-dimensionality, such as flow in
corners, the fluid motions are controlled by the momentum flux and
mean-shear. Characteristic scales consistent with the previous
argument are $\mathcal{U}_2=\sqrt{\tau_w/\rho}$ and
$\mathcal{L}_2=\mathcal{U}_2/S$~\citep{Lozano2019}, where
$S=||\nabla\langle\boldsymbol{u}\rangle||$. In separated flows, the
mean momentum equation is balanced by the viscous terms and pressure
gradient, which gives $\mathcal{U}_3=[(\partial \langle p
  \rangle/\partial s)\nu/\rho]^{1/3}$ and
$\mathcal{L}_3=\nu/\mathcal{U}_3$~\citep{Stratford1959}, where
$\langle p\rangle$ is mean pressure and $s$ the mean streamwise
direction. The values of $\partial{\langle p\rangle}/\partial s$ and
$S$ are evaluated using the discrete seven-point stencil and the wall
stress vector $\tau_w$ is obtained from the previous time step.

Finally, the input to the wall model comprises the non-dimensional
groups formed by the dimensional set
\begin{equation}
  \left\{ \tilde{u}_i^{\boldsymbol{0}}, \tilde{S}_{ij},
  \tilde{\Omega}_{ij} ; \mathcal{U}_k , \mathcal{L}_k ; U_\infty,
  \rho, \nu, \Delta, h  : \ i,j,k=1,2,3\right\},
\end{equation} 
which constitutes a collection of local Reynolds numbers and velocity
differences non-dimensionalized by the characteristic scales. Examples
of non-dimensional input variables can be found in Figure
\ref{fig:model_sketch}.

The output variables are divided into two sets: the wall stress vector
expressed as $10^3 \times \boldsymbol{\tau_w}/(\rho U_\infty^2)$ and
the information about the model flow classification and confidence on
the solution. The last two outputs are explained in more detail in the
next subsection. 
  
\subsection{Neural network architecture and training}

The predictor-classifier structure is outlined in
Figure~\ref{fig:model_sketch}.  The predictor is a deep feed-forward
neural network with 4 hidden layers and 30 neurons per layer.  Both
the inputs and outputs of the predictor are standardized. The
classifier is a Bayesian neural network with 3 hidden layers and 15
neurons per layer. The activation functions selected for the hidden
layers are hyperbolic tangent sigmoid transfer functions and rectified
linear activation transfer functions.  The present layout was found to
give a fair compromise between neural network complexity and
predictive capabilities. Nonetheless, future versions of the model
might be accompanied by a systematic optimization of the neural
network architecture.

Note that we would not even need a neural network to perform the steps
above. The classification step can be conducted by means of other
clustering techniques (such as k-means analysis), and the wall-stress
prediction might be attained by switching between different analytic
models (akin to the equilibrium wall model in Eq.~(\ref{eq:algwm}))
according to the flow classification. Nonetheless, neural networks
have shown excellent performance in classification tasks, and they are
a convenient tool to build non-linear interpolations among datasets.

The training is performed using the following DNS databases: turbulent
channel flows at $Re_\tau\approx350, 550, 1000, 2000$ and
$4200$~\citep{Lozano2014a}, turbulent ducts at $Re_\tau\approx 180$
and $390$ and aspect ratios equal to 1, 3, 6 and
10~\citep{Vinuesa2014}, and turbulent boundary layers with a separation
bubble via blowing and suction at $Re_\tau\approx 180$ following the
setup from \cite{Na1998}.  The input signal is obtained by sampling
the DNS data for various fictitious (isotropic) grid resolutions
ranging from $\Delta/\delta = 0.01$ to $1$ and at multiple wall-normal
locations from $0.01\delta$ to $\delta$, where $\delta$ is the
boundary-layer thickness (or channel half-height). The DNS velocity is
averaged within a sphere of radius $\Delta$ centered at the location
of $\boldsymbol{x}^{\boldsymbol{0}}$ (similarly for
$\boldsymbol{x}^{\pm \boldsymbol{e}_i}$). The output is taken as the
wall stress averaged over the surface area $\Delta^2$. The training
set is also augmented by performing arbitrary rotations of the frame
of reference.  The neural network is trained using Bayesian
regularization backpropagation by randomly dividing the training date
into two groups, the training set (80\% of the data) and test set
(20\% of the data).

\section{Model validation: Results of WMLES with SCMv1}\label{sec:validation}

\begin{figure}
  \centering
  \vspace{1cm}
\includegraphics[width=1\textwidth]{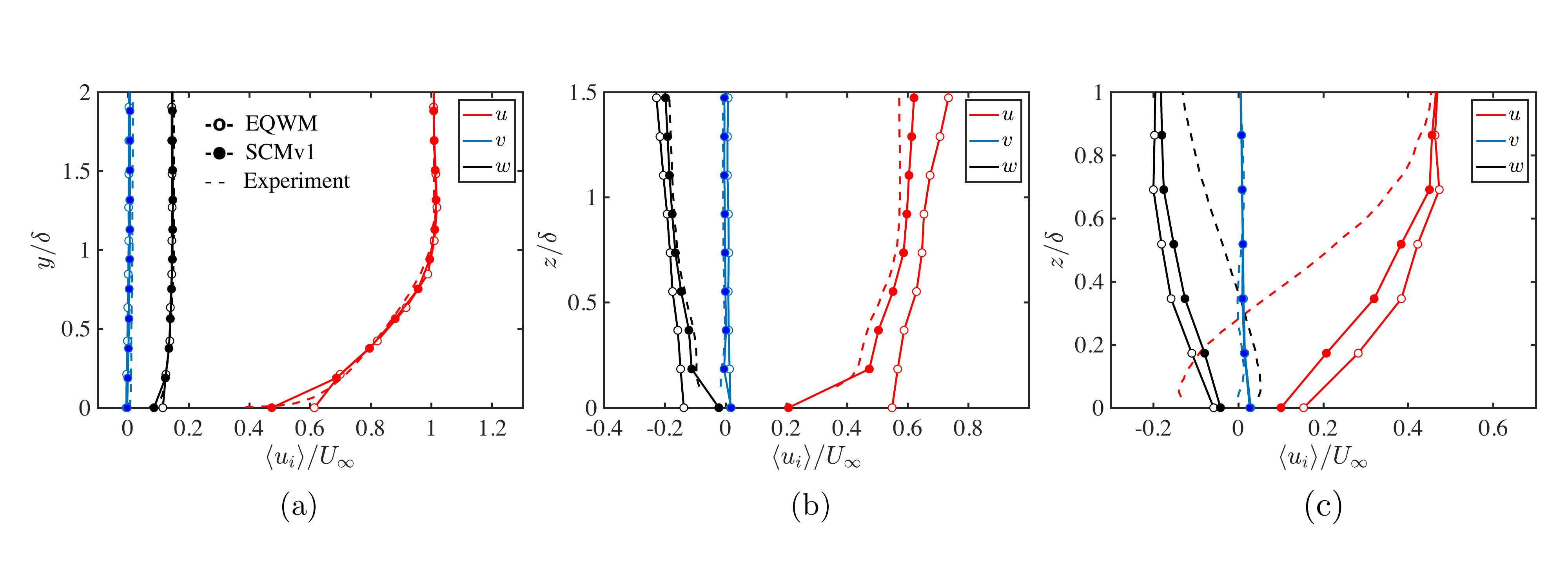}
\caption{ The mean velocity profiles at (a) upstream region of the
  fuselage at $x=1168.4$~mm and $z=0$~mm, (b) wing-body juncture at
  $x=2747.6$~mm and $y=239.1$~mm, and (c) wing-body juncture close to
  the trailing edge at $x=2922.6$~mm and $y=239.1$~mm. Solid lines with
  open symbols are for WMLES with EQWM and closed symbols for WMLES
  with SCMv1. Dashed lines are experiments. Colors denote
  different velocity components. The distances $y$ and $z$ are
  normalized by the local boundary-layer thickness $\delta$ at each
  location.\label{fig:results_SCM}}
\end{figure}

We validate the model in the NASA Juncture Flow Experiment. The
problem setup is identical to the one described in Section
\ref{sec:methods} replacing the equilibrium wall model by the
SCMv1. The prediction of the mean velocity profiles is presented in
Figure~\ref{fig:results_SCM} and compared with WMLES with EQWM for the
three locations under consideration.  Table~\ref{table:results_SCM}
contains information about the flow classification and confidence on
the solution at each location.  The relative error on the prediction,
quantified as $||\boldsymbol{u}^{\mathrm{exp}} -
\boldsymbol{u}||/||\boldsymbol{u}^{\mathrm{exp}}||$, is included in the
last rows of Table~\ref{table:results_SCM} for both SCMv1 and EQWM.
%
\begin{table}
  \centering
\begin{tabular}{llrrrl}
                                   &                                                                                                            & Location (a)                                                & Location (b)                                                & Location (c)                                                 &  \\ \hline
\multicolumn{1}{c}{Classification} & \multicolumn{1}{c}{\begin{tabular}[c]{@{}c@{}}Flat plate flow\\ Corner flow\\ Separated flow\end{tabular}} & \begin{tabular}[c]{@{}r@{}}91\% \\  9\% \\ 0\%\end{tabular} & \begin{tabular}[c]{@{}r@{}}27\% \\ 73\% \\ 0\%\end{tabular} & \begin{tabular}[c]{@{}r@{}}20\% \\ 52\% \\ 28\%\end{tabular} &  \\ \hline
\multicolumn{1}{c}{Confidence}     & \multicolumn{1}{c}{}                                                                                       & 96\%                                                        & 92\%                                                        & 20\%                                                         &  \\
&                                                                                                            & \multicolumn{1}{l}{}                                        & \multicolumn{1}{l}{}                                        & \multicolumn{1}{l}{}                                         & \\ \hline 
Error SCMv1    &  & 1.4\%  &  7.4\%  & 78.9\% \\
Error EQWM     &  & 2.0\%  & 24.5\%  & 98.7\%  
\end{tabular} 
\caption{The flow classification, confidence on the solution, and
  error in the mean velocity profiles by SCMv1. The errors in the last row
  are for WMLES with EQWM. The locations are (a) upstream region of the
  fuselage, (b) wing-body juncture, and (c) wing-body juncture close
  to the trailing edge.\label{table:results_SCM}}
\end{table}

Figure~\ref{fig:results_SCM}(a) shows that the predictions for SCMv1
and EQWM at the fuselage location are roughly identical. In both
cases, the wall stress is within $2\%$ accuracy as reported in
previous investigations~\citep{Lozano_brief_2020_1}. The flow is
identified by SCMv1 as a flat-plate turbulence with $96\%$
confidence.  The outcome is expected as SCMv1 was trained in turbulent
channel flows and we have argued that the boundary layer at the
fuselage resembles a ZPGTBL.  The most notable improvement is found at
the juncture location (Figure~\ref{fig:results_SCM}b): SCMv1 provides
an augmented value of the wall stress, which alleviates the
overprediction of $u$ and $w$ using EQWM.  The new wall-stress
prediction of the mean velocity profiles is to within $7.4$\% error,
compared with the 24.5\% error using EQWM.  The flow is classified as
predominantly corner flow with some traces of flat-plate
turbulence. The confidence in the prediction is also above $90\%$.

Finally, the performance of SCMv1 in the separated region
(Figure~\ref{fig:results_SCM}c) is the poorest of all three locations,
but also the most interesting.  There is a moderate improvement on the
mean velocity prediction by SCMv1 with respect to EQWM, but this is
still far from being satisfactory and errors remain above $78.9\%$.
Interestingly, SCMv1 classifies the flow as a mix of flat plate
turbulence, corner flow, and separated flow. Moreover, despite the
erroneous prediction by SCMv1, the model prompts a warning about its
poor performance, which is evidenced by the low confidence on the
wall-stress prediction ($\sim$20\%).  The deficient performance of
both wall models is not surprising if we note that the separation zone
has a wall-normal thickness of 0.3$\delta$, whereas the WMLES grid
size is $\Delta\approx 0.2\delta$. Thus, there is only one grid point
across the separation bubble. Despite the fact that the model was
trained in separated flows, numerical errors and SGS model errors
dominate the LES solution in this case. These errors hinder the
capability of SCMv1 to classify the flow. Nonetheless, the ability of
SCMv1 to assess the confidence on the prediction is a competitive
advantage with respect to traditional wall models.
%

\section{Conclusions and outlook}\label{sec:conclusions}

The prediction of aircraft aerodynamic quantities of interest remains
among the most pressing challenges for computational fluid dynamics,
and it has been highlighted as a Critical Flow Phenomena in the NASA
CFD Vision 2030~\citep{Slotnick2014}. The aircraft aerodynamics are
inherently turbulent with mean-flow three-dimensionality, often
accompanied by laminar-to-turbulent transition, flow separation,
secondary flow motions at corners, and shock wave formation, to name a
few. However, the most widespread wall models are built upon the
assumption of statistically-in-equilibrium wall-bounded turbulence and
do not faithfully account for the wide variety of flow conditions
described above.  This raises the question of how to devise models
capable of accounting for such a vast and rich collection of flow
physics in a feasible manner.

In this preliminary work, we have proposed tackling the wall-modeling
challenge by devising the flow as a collection of building blocks,
whose information enables the prediction of the stress as the wall. We
refer to the wall model as SCMv1 (self-critical model version 0.1).
The model relies on the assumption that simple canonical flows (such
as turbulent channel flows, boundary layers, pipes, ducts, speed
bumps, etc) contain the essential flow physics to devise accurate
models. Three types of building block units were used to train the
model, namely, turbulent channel flows, turbulent ducts and turbulent
boundary layers with separation. This limited training set will be
extended in future versions of the model.  The approach is implemented
using two interconnected artificial neural networks: a classifier,
which identifies the contribution of each building block in the flow;
and a predictor, which estimates the wall stress via non-linear
combinations of building-block units.  The output of the model is
accompanied by the confidence in the prediction. The latter value aids
the detection of areas where the model underperforms, such as flow
regions that are not representative of the building blocks used to
train the model.  This is the self-critical component of the SCM and
is considered a key step for developing reliable models. For example,
the present model will provide a low confidence value in the presence
of a flow that it has never seen before (e.g., a shock wave), or when
the input velocity is outside of the boundary layer.

The model was validated in an unseen case representative of external
aerodynamic applications: the NASA Juncture Flow Experiment.  The case
is a generic full-span wing-fuselage body at Reynolds number
$Re=2.4\times 10^6$.  We have characterized the WMLES errors in the
prediction of mean velocity profiles with SCMv1 and a widely used
equilibrium wall model. Three different locations over the aircraft
have been considered: the upstream region of the fuselage, the
wing-body juncture, and the wing-body juncture close to the trailing
edge. The last two locations are characterized by strong mean-flow
three-dimensionality and separation. We have shown that SCMv1
outperforms the EQWM in the three locations investigated. However, the
most remarkable result is not the higher accuracy of SCM which, due to
the larger number of degrees of freedom in the model, is deemed to
outperform the EQWM.  Instead, we remark on (1) the success of the
model in providing confidence levels in the wall-stress prediction and
(2) its potential to account for new flow physics by including
additional building block units. The promising performance of SCMv1
presented here is still limited to a few observations and thus should
be taken with caution.  Further investigation is needed to
systematically characterize errors at multiple grid resolutions and
flow configurations.


Several outstanding issues remain to be solved, such as the
identification of meaningful building-block flows, the minimum number
of blocks required to make accurate predictions, and their
characteristic scales. The choice of input and output variables and
their non-dimensionalization is also of paramount importance in
obtaining a successful model. Here, we have used instantaneous data as
input information but time-varying signals are probably needed to
predict strongly unsteady effects. Another outstanding issue is the
necessity of high-quality wind-tunnel experiments as a proxy for
evaluating the accuracy of WMLES in real-world external aerodynamic
applications.  Useful measurements include pointwise time-averaged
velocities and Reynolds stress profiles along with surface pressure
and friction coefficients. The scarcity to date of granular
experimental quantities hinders our ability to assess and improve the
performance of WMLES in more realistic scenarios.

We have argued that truly revolutionary improvements in WMLES will
encompass advancements in numerics, grid generation, and wall/SGS
modeling.  Here, we have focused on wall-modeling and much work
remains to be done on other fronts.  There is obviously a data science
component to the problem too, such as the need for efficient and
reliable machine-learning techniques for data classification and
regression. However, the main emphasis of this work has been on the
physical understanding of the problem rather than on the details of
neural network architecture at hand.  The comment in the introduction
about the shallowness of the statement ``LES modeling will be solved
by machine learning'' is just a reminder that problems are not solved
by tools but by people.

\section*{Acknowledgments}

A.L.-D. acknowledges the support of NASA under grant No. NNX15AU93A. 

\bibliographystyle{ctr}

\end{document}